\documentclass{article}

\usepackage{multirow}
\usepackage{graphicx} 
\usepackage{amsmath}
\usepackage{enumitem}
\usepackage[preprint]{neurips_2025}

\usepackage[utf8]{inputenc} 
\usepackage[T1]{fontenc}    
\usepackage{hyperref}       
\usepackage{url}            
\usepackage{booktabs}       
\usepackage{amsfonts}       
\usepackage{nicefrac}       
\usepackage{microtype}      
\usepackage{xcolor}         

\title{Act-With-Think: Chunk Auto-Regressive Modeling for Generative Recommendation}

\title{Act-With-Think: Chunk Auto-Regressive Modeling for Generative Recommendation}

\author{%
  \textbf{Yifan Wang}$^{1}$\thanks{Equal contribution}, 
  \textbf{Weinan Gan}$^{2*}$, \textbf{Longtao Xiao}$^{1}$, \textbf{Jieming Zhu}$^{2\dag}$, \textbf{Heng Chang}$^{3}$, \\ \textbf{Haozhao Wang}$^{1}$, \textbf{Rui Zhang}$^{1}$, \textbf{Zhenhua Dong}$^{2}$, \textbf{Ruiming Tang}$^{2}$, \textbf{Ruixuan Li}$^{1}$\thanks{Corresponding authors}\\
  $^{1}$ Huazhong University of Science and Technology, 
  $^{2}$ Noah’s Ark Lab, Huawei \\
  $^{3}$ Huawei Technologies Co., Ltd
  \\d202381481@hust.edu.cn, ganweinan1@huawei.com
}

\begin{document}
\bibliographystyle{plain}
\maketitle

\begin{abstract}
Generative recommendation (GR) typically encodes behavioral or semantic aspects of item information into discrete tokens, leveraging the standard autoregressive (AR) generation paradigm to make predictions. 
However, existing methods tend to overlook their intrinsic relationship, that is, the semantic usually provides some reasonable explainability ``\textbf{why}'' for the behavior ``\textbf{what}'', which may constrain the full potential of GR. 
To this end, we present Chunk AutoRegressive Modeling (CAR), a new generation paradigm following the decision pattern that users usually think semantic aspects of items (e.g. brand) and then take actions on target items (e.g. purchase). Our CAR, for the \textit{first time}, incorporates semantics (SIDs) and behavior (UID) into a single autoregressive transformer from an ``act-with-think'' dual perspective via chunk-level autoregression. 
Specifically, CAR packs SIDs and UID into a conceptual chunk for item unified representation, allowing each decoding step to make a holistic prediction. 
Experiments show that our CAR significantly outperforms existing methods based on traditional AR, improving Recall@5 by 7.93\% to 22.30\%. 
Furthermore, we verify the scaling effect between model performance and SIDs bit number, demonstrating that CAR preliminary emulates a kind of slow-thinking style mechanism akin to the reasoning processes observed in large language models (LLMs). 
\end{abstract}

\section{Introduction}
In recent years, generative models, particularly large language models based on autoregressive generation, have achieved significant success~\citep{llm1,llm2,llm3}. 
This successful experience inspired the exploration of technical paradigms from matching to generating in the recommendation system~\citep{gr1,shuangta,matrix,bert4rec}. An early example of generative recommendation is TIGER~\citep{tiger}, which transforms textual item descriptions into Semantic IDs (SIDs) and predicts user interests via transformer-based sequence-to-sequence modeling~\citep{sentencet5,t5}. 
SIDs are tokens derived from discretized text embeddings, which reflect the semantic information of items and enable the model to recommend items that are semantically similar to the user’s interaction history~\citep{sid,km}.
For instance, the SIDs tuple of an item is (5, 23, 55), where 5 represents ``shoes'', 23 corresponds to ``orange'', and 55 denotes ``brandX''. If a user has frequently browsed shoe-related items, the system can infer a continued interest in that category based on SIDs, and recommend items whose SIDs are semantically similar to ``shoes''.

However, models based solely on SIDs overlook collaborative information modeling—the semantic information(SIDs) between co-occurring items may differ significantly, making it difficult to recommend co-occurring items based on semantic relevance~\citep{actionpiece,eager}. Collaborative information is typically captured by unique IDs (UIDs), which determine what is recommended, but lack semantics, making it difficult to explain why the recommendation is made~\citep{sasrec,s3}. Several recent efforts have attempted to integrate collaborative information into generative recommendation frameworks~\citep{eager,cobra}. While differing in technical implementation, most of these approaches follow a similar conceptual pattern: treating semantic and collaborative information as independent feature spaces. This independence assumption neglects their inherent interdependence in that a UID is recommended because it is semantically aligned with the user’s historical interactions. Therefore, decoupling semantic information (SIDs) from behavioral information (UID) hinders the model’s ability to accurately model user interests, and to establish an internal logic chain for explaining recommendations.

Inspired by the concept of slow-thinking \citep{slow1, slow2, slow3}, which has proven effective in improving LLM reasoning on complex tasks via deeper intermediate computation, we propose Chunk Autoregressive Modeling(CAR). By introducing an ``act-with-think'' perspective, CAR brings this slow-thinking style machanism into generative recommendation, aiming to preserve generation efficiency while more effectively integrating semantic information with behavioral patterns, thereby enhancing the model’s performance.
Specifically, we treat UIDs as the user's action to capture collaborative information, and the associated set of SIDs as the semantic thoughts that drive this action to capture semantic information. This SIDs–UID tuple is treated as a composite chunk, upon which we build a chunk-level autoregressive modeling framework to reflect the ``act-with-think'' decision process. Chunk-level autoregressive modeling, by increasing intermediate computational steps akin to the slow-thinking mechanism in NLP, enables the model to jointly learn semantic information and behavioral patterns~\citep{var,deepseek}. This design reflects the way users typically make decisions—by first considering semantic aspects of items (such as category, brand, or color), and then taking actions (such as clicks or purchases) based on a synthesis of this semantic information and their historical behavior patterns. This perspective enables the model to more effectively align with the cognitive process behind user interactions, rather than treating semantic and behavioral signals as isolated sources of information.
To sum up, our core contributions are summarized as follows:
\begin{itemize}
\item We propose CAR, a novel chunk-level autoregressive paradigm for generative recommendation.
CAR models user's history as a sequence of SIDs-UID chunks, enabling joint learning of user actions and their semantic information behind these actions in a unified framework.
\item We introduce an ``act-with-think'' perspective that models the interdependence between semantic and collaborative information. This perspective facilitates a seamless and intrinsic integration of semantic and collaborative information, while embedding a slow-thinking style mechanism into the generative recommendation process to enhance performance.
\item We demonstrate the effectiveness of CAR through extensive experiments on three public datasets.
CAR achieves 7.93\%–22.30\% improvements in Recall@5 over strong baselines.
\end{itemize}

\begin{figure}
  \centering
  \includegraphics[width=\textwidth]{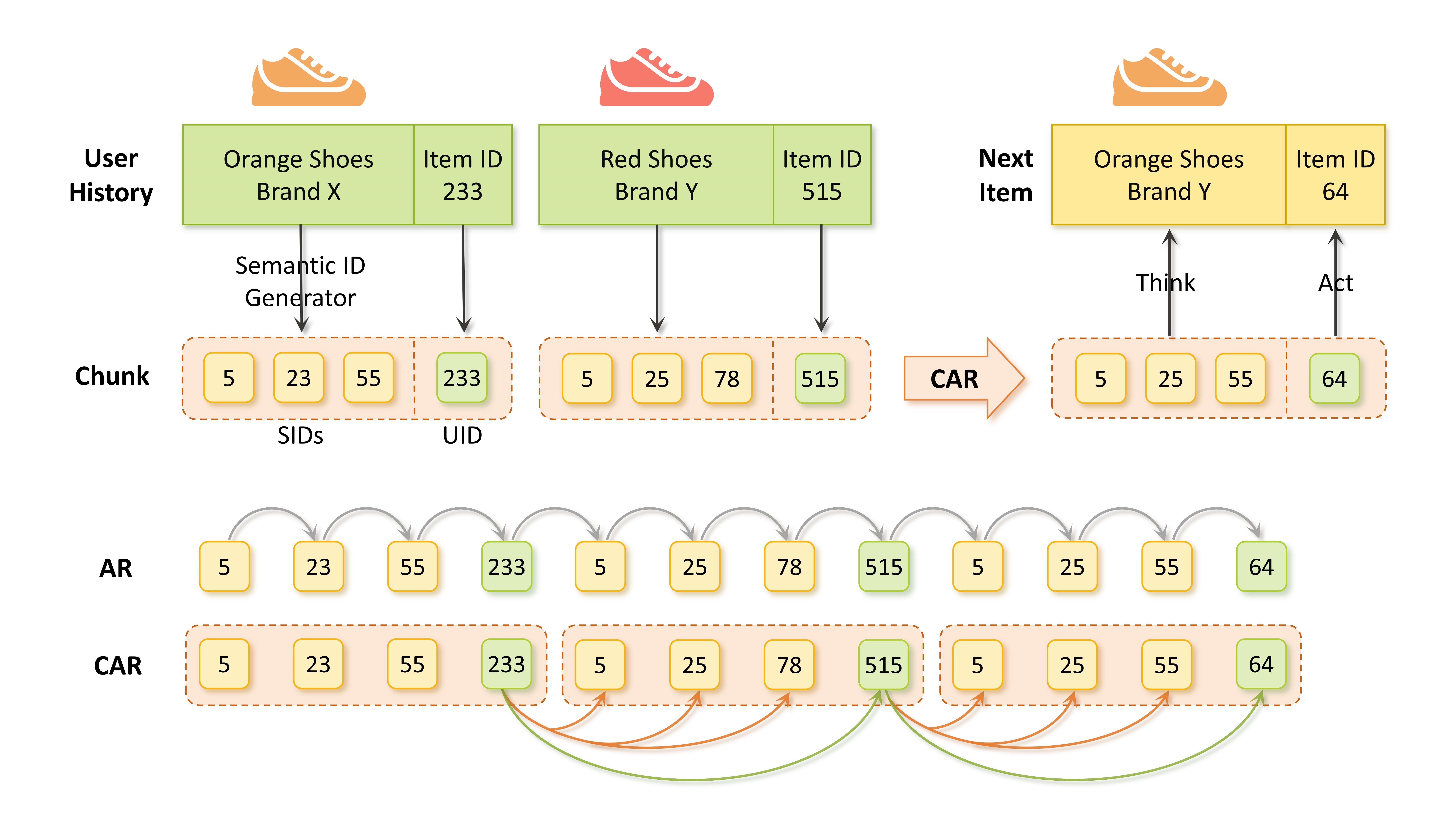}
  \caption{An overview of the Chunk AutoRegressive Modeling framework. First, each item is transformed into an ``act-with-think'' chunk that includes SIDs and an UID, serving as the basic modeling unit. Then, autoregressive modeling is performed at the chunk level. The bottom part of the figure further contrasts the modeling approaches of standard autoregressive modeling (AR) and our proposed method, CAR.}
  \label{overview}
\end{figure}

\section{Related Work}

\paragraph{Generative Recommendation.}
Generative recommendation is an emerging recommendation paradigm, and its core process includes two stages: discrete semantic tokenization and autoregressive sequence generation~\citep{gr1,dsi,nci}. TIGER~\citep{tiger}, as a pioneering approach in the field, achieves sequence generation by encoding item textual information into semantic IDs and implementing sequence generation based on the T5 architecture~\citep{sentencet5,t5}. EAGER~\citep{eager} proposed the dual-stream generation framework, which achieves information fusion through the parallel prediction of semantic encoding and collaborative encoding. Nonetheless, the architectural complexity and insufficient interaction during decoding hinder the deep fusion of collaborative semantics. COBRA~\citep{cobra}, on the other hand, attempts to combine discrete semantic encoding with dense embeddings.Through joint training, the embedding can carry both collaborative and semantic information simultaneously.  However, the intrinsic heterogeneity between these types of information makes it challenging for a single vector to effectively represent both characteristics. In contrast, our approach adopts an ``act-with-think'' perspective, reconstructing the autoregressive generation process to achieve a natural fusion of collaborative and semantic information, while preserving their relative independence.

\paragraph{Slow Thinking Reasoning Systems.}
The core idea of slow thinking lies in enhancing decision quality through deep reasoning and systematic analysis, which is primarily achieved by increasing the number of intermediate computational steps~\citep{slow1,slow2,slow3}. Fundamentally, it constructs a complete logical chain prior to action, thereby reducing cognitive biases and errors. OpenAI’s o1~\citep{o1} model adopts this ``slow thinking'' paradigm by generating logically coherent solutions through multiple rounds of introspective reasoning. Similarly, DeepSeek-R1~\citep{r1} implements a three-stage training strategy—imitation, exploration, and self-improvement—enabling the model to autonomously generate reasoning chains and dynamically refine its inference paths. These slow thinking frameworks demonstrate that deliberate reasoning leads to more comprehensive and reliable solutions~\citep{deepseekthink,thinkrc,slowrc}. Inspired by this, we propose a novel ``act-with-think'' perspective to re-examine the integration of collaborative and semantic information in generative recommendation systems.

\paragraph{AutoRegressive Modeling.}
Autoregressive models, which operate via conditional probability mechanisms to predict subsequent elements in a sequence based on prior inputs, have demonstrated remarkable performance in the field of Natural Language Processing (NLP)~\citep{llm1,llm2,llm3}. However, as deep learning technologies have evolved, the traditional step-by-step prediction paradigm has increasingly revealed limitations in both efficiency and modeling capacity. To address these challenges, the Multi-Token Prediction (MTP) approach~\citep{deepseek} was introduced. By enabling the simultaneous prediction of multiple tokens at each training step, MTP not only significantly improves training efficiency but also enhances the model’s ability to capture long-range dependencies.
In the domain of computer vision, Visual Autoregressive Modeling (VAR)~\citep{var} further extends the applicability of autoregressive techniques by incorporating a Next-Scale Prediction strategy. This approach generates images progressively from low to high resolutions, leading to notable improvements in both image quality and generation speed. As a result, autoregressive models have, for the first time, outperformed diffusion-based methods in image synthesis tasks, exhibiting strong scalability and zero-shot generalization capabilities\citep{ControlAR}.
Inspired by the aforementioned works, we reconstruct the autoregressive modeling for generative recommendation based on ``act-with-think'' chunks.

\section{Chunk-level AutoRegressive Modeling (CAR)}

Our proposed framework consists of three components:
\begin{enumerate}

    \item \textit{Constructing ``act-with-think'' chunk.}
    We first encode the textual descriptions of items into embedding vectors and discretize the embeddings into non-unique SIDs using residual KMeans, where the SIDs capture the semantic ``thought'' of the items. Then, we concatenate these SIDs with the corresponding UID, which denote the ``act'' operations, to construct the ``act-with-think'' chunk.
    
    \item  \textit{Act-Think Co-Generation. }
    To model the ``act-with-think'' relationship, we predict the ``act'' UID and the ``think'' SIDs simultaneously. This is achieved by using the last token of the current chunk to generate all tokens of the next chunk at once.
    
    \item \textit{Progressive Act-Think Context Fusion.} 
    To further enhance the ``act-with-think'' representation, we leverage the co-occurrence patterns inherent in chunks by progressively integrating prefix information into the input, thereby enriching the contextual representation of the current token~\citep{n-gram}.

\end{enumerate}

\begin{figure}
  \centering
  \includegraphics[width=\textwidth]{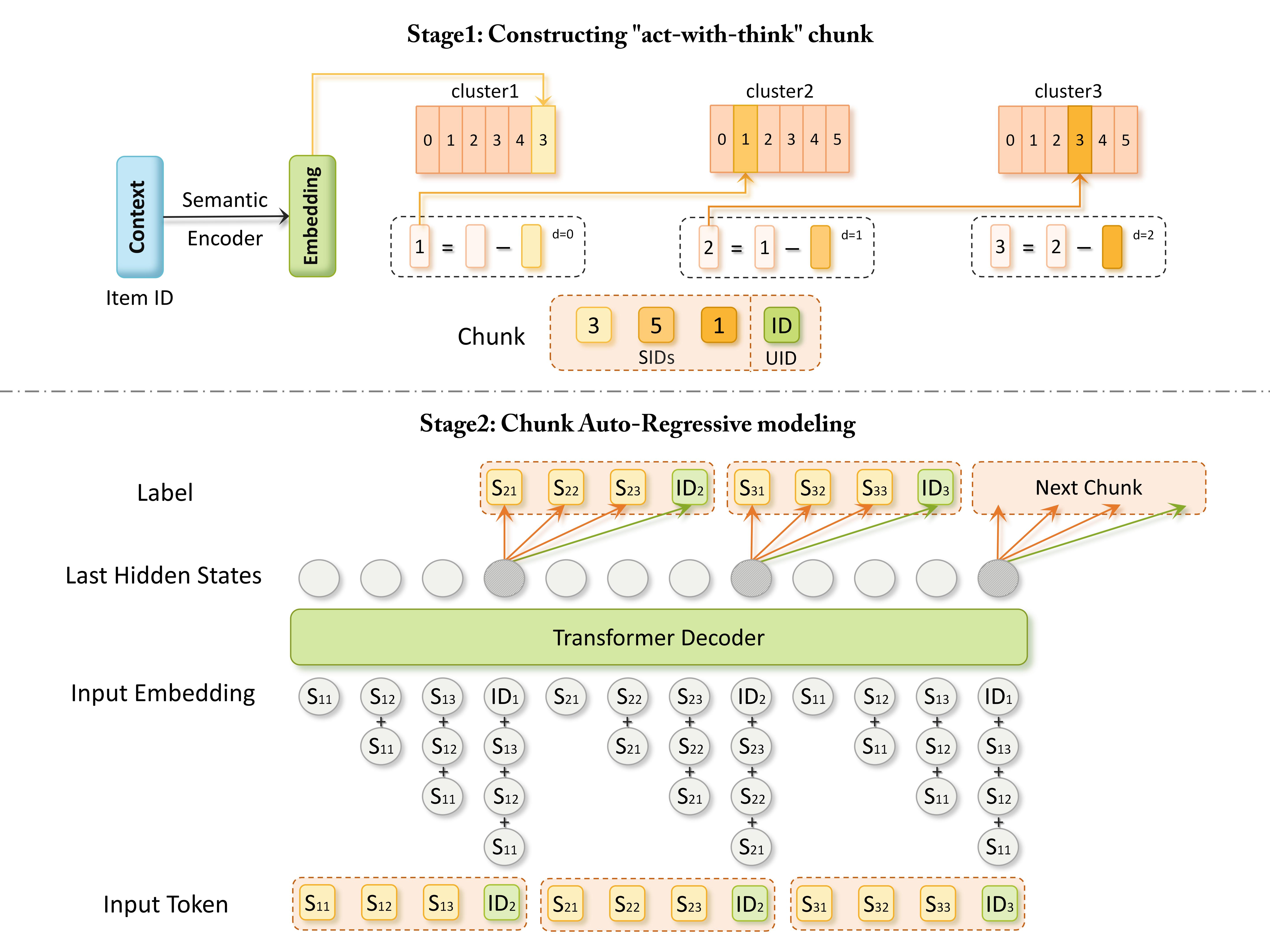}
  \caption{CAR consists of two independent training stages.
Stage 1: Semantic IDs (SIDs) are generated using residual KMeans and combined with Unique ID (UID) to construct the basic modeling unit, the ``act-with-think'' chunk.
Stage 2: On the input side, token representations within each chunk are progressively enhanced via prefix-based embedding augmentation. During training, autoregressive modeling is performed at the chunk level.}
  \label{CAR}
\end{figure}

\subsection{Constructing ``Act-With-Think'' Chunk}
This section describes the construction process of our proposed ``act-with-think'' chunk, as illustrated in Fig.~\ref{CAR}. First, we utilize a pre-trained text encoder (Sentence-T5~\citep{sentencet5}) to encode item content features (e.g., title and description) into high-dimensional semantic embedding vectors. Next, we perform hierarchical discretization of these embeddings using a multi-level residual KMeans algorithm, generating non-unique SIDs that capture shared semantic information across items—representing the ``think'' dimension. Finally, we concatenate these SIDs with the UID of each item (representing the ``act'' dimension), forming an ``act-with-think'' chunk that integrates semantic information with discrete item behavior.

\paragraph{Residual KMeans for Semantic IDs.}
Unlike the RQVAE-based approach used in TIGER \citep{tiger}—which forces codebook uniformity to generate unique hierarchical SIDs per item—our method avoids semantic distortion. RQVAE~\citep{rqvae} also requires joint training of neural encoder-decoder networks and quantization codebooks, leading to potential codebook collapse and training instability. In contrast, we propose a parameter-free residual KMeans quantization strategy to generate non-unique hierarchical SIDs directly from the original embeddings. At each level, k-means clustering is applied to the residuals of the previous quantization step, enabling the extraction of inherent hierarchical structure in the embedding distribution. This approach is data-driven and model-free, resulting in more faithful and flexible semantic representations that better reflect the shared nature of semantic information.

\subsection{Act-Think Co-Generation}

Based on a user's historical interaction sequence \textit{H}: \((\text{item}_1, \ldots, \text{item}_n)\), we model the synchronized prediction of ``act'' and ``think'' as illustrated in Fig.~\ref{overview}. Specifically, each item is represented as a composite chunk \((S_{1}, S_{2}, \ldots, \text{UID})\), where the semantic IDs \((S_1, S_2, \ldots)\) captures semantic information, while the unique ID (\(\text{UID}\)) preserves item uniqueness.

Traditional autoregressive methods typically adopt a chain decomposition for multi-granularity prediction~\citep{attention,neural}:

\begin{equation}
P(S_1, S_2, \text{UID} \mid H) = P(S_1 \mid H) \cdot P(S_2 \mid H, S_1) \cdot P(\text{UID} \mid H, S_1, S_2)
\end{equation}

This formulation implicitly assumes that the prediction of UID depends on the generated SIDs. However, semantic and collaborative information capture different aspects of item characteristics~\citep{eager}. While these two dimensions are often correlated, enforcing a strict dependency between them may lead to suboptimal representations. In other words, our goal is to model the dimensions of ``action'' and ``think'' in a relatively independent and parallel manner, while maintaining a strong connection between them.

To address this issue, we advocate for a more flexible modeling strategy that treats ``think'' (SIDs) and ``act'' (UID) as relatively independent yet interrelated processes. Our goal is to model these dimensions in parallel, preserving their mutual relevance without imposing a rigid hierarchical dependency.
To this end, we propose the ``act-with-think'' modeling strategy, which reformulates the joint prediction task as:

\begin{equation}
P(S_1, S_2, \text{UID} \mid H) = P(S_1 \mid H) \cdot P(S_2 \mid H) \cdot P(\text{UID} \mid H)
\end{equation}

This factorization explicitly decouples the prediction of collaborative and semantic information, as well as removes the hierarchical dependency across semantic levels. By doing so, it not only enables more flexible representations of item characteristics, but also simplifies the prediction process. In particular, it eliminates the need for sequential decoding—such as beam search—thus avoiding the associated combinatorial complexity.Furthermore, this formulation facilitates chunk-level autoregressive modeling: by using the final token of the current chunk, the model can simultaneously predict all tokens in the next chunk in parallel. This design significantly improves inference efficiency while preserving the temporal structure of the autoregressive framework.

To align with the joint prediction objective of ``act-with-think'', we design a dual-branch loss function that separately supervises semantic abstraction and item-level discrimination:
\begin{itemize}[left=0pt]
    \item \textit{Think branch}: To guide the generation of SIDs at each level, we define the think loss as:
    \begin{equation}
  \mathcal{L}_{\text{think}} = \frac{1}{n} \sum_{i=1}^{n} \log P(S_i \mid H)
    \end{equation}  
    \item \textit{Act branch}: To ensure accurate prediction of the UID, we introduce the act loss as:
    \begin{equation}
    \mathcal{L}_{\text{act}} = \log P(\text{UID} \mid H)
    \end{equation}
\end{itemize}

The overall training objective combines both losses:

\begin{equation}
\mathcal{L} = \alpha \cdot \mathcal{L}_{\text{think}} + \mathcal{L}_{\text{act}}
\end{equation}

Here, \(\alpha\) is a hyperparameter that controls the trade-off between capturing semantic commonality across items and emphasizing item-specific accuracy. This dual-branch structure enables the model to capture high-level semantic information through the think branch while delivering precise item recommendation through the act branch.

\subsection{Progressive Act-Think Context Fusion}

To enhance intralayer semantic inheritance within SIDs and cross-modal coordination between semantic and collaborative information, we propose a Progressive Act-Think Context Fusion mechanism (as illustrated in Fig.~\ref{CAR}). This mechanism constructs the representation of each token within a chunk by fusing think-level semantic information (e.g., \(S_1, S_2\)) from its historical prefix.

Concretely, for a token at position \(k\), its representation is obtained by aggregating its current token embedding with the semantic embeddings of all think levels from the previous \(k-1\) positions. This results in a cross-level hybrid context representation that enables the model to simultaneously capture two key patterns:

1. \textit{Vertical Semantic Inheritance}: Higher semantic levels (e.g., \(S_2\)) inherit and refine coarse-grained semantics from shallower levels (e.g., \(S_1\)), supporting a progressive abstraction of semantic information across the hierarchy.

2. \textit{Horizontal Cross-Modal Coordination}: By integrating act (UID) and think (SIDs) representations into a shared contextual space, the model jointly encodes both semantic generality and entity-level specificity. This coordinated fusion enhances the expressiveness of item representations and improves recommendation accuracy.

\section{Experiments}

\paragraph{Datasets}
In this experiment, we use the Amazon Product Review dataset (spanning May 1996 – July 2014)~\citep{amazon} and select three product categories to construct the benchmark dataset: ``Beauty'', ``Sports \& Outdoors'', and ``Toys \& Games''. We follow the 5-core filtering standard used in TIGER~\citep{tiger}, ensuring that each user/item has at least 5 interaction records. Table~\ref{datasets} presents key statistics of the processed dataset.

\begin{table}
  \caption{Statistics of the datasets}
  \label{datasets}
  \centering
  \begin{tabular}{ccccc}
    \toprule
    \multirow{2}{*}{Dataset} &\multirow{2}{*}{Users} &\multirow{2}{*}{Items} &\multicolumn{2}{c}{Sequence Length}                \\
    \cmidrule(lr){4-5}
     & & &Mean &Median \\
    \midrule
    Beauty &22363 &12101 &8.87 &6     \\
    Sports and Outdoors &35598 &18357 &8.32 &6      \\
    Toys and Games &19412 &11924 &8.63 &6  \\
    \bottomrule
  \end{tabular}
\end{table}

\paragraph{Evaluation Metrics}
We evaluate recommendation performance using top-k recall (Recall@K) and normalized discounted cumulative gain (NDCG@K) for K = 5, 10.
For each item sequence, the last item is used for testing, the second-last item is used for validation, and the remaining items are used for training~\citep{sasrec}.
During training, we limit the number of historical items per user to 20.

\paragraph{Implementation Details}
The model is built using the ``GPT2LMHeadModel'' from the Hugging Face Transformers library~\footnote{https://github.com/huggingface/transformers}, configured with one transformer layer and eight self-attention heads. The hidden size of the MLP is set to 1024, and the input embedding dimension is 128. A dropout rate of 0.1 is applied to prevent overfitting. The model is trained with a learning rate of 1e-5 and a batch size of 256.

\subsection{Performance Comparison}

The baseline methods for comparison are categorized into three groups:
\begin{enumerate}
    \item \textit{Traditional sequential methods:}\begin{itemize}
        \item \textbf{Caser}~\citep{caser}: Caser models user behaviors as an ``image'' of past interactions and leverages convolutional filters to capture both union-level and sequential patterns.
        \item \textbf{HGN}~\citep{hgn}: HGN introduces a gated mechanism to dynamically fuse long- and short-term user preferences for more personalized sequence modeling.
        \item \textbf{SASRec}~\citep{sasrec}: SASRec applies self-attention to capture long-range dependencies in user behavior sequences, enabling flexible and interpretable modeling of preferences.
        \item \textbf{S3Rec}~\citep{s3}: S3Rec incorporates self-supervised learning objectives to enhance item representations and user modeling from sparse sequential data.
    \end{itemize} 
    \item \textit{Generative methods:} \begin{itemize}
        \item \textbf{TIGER}~\citep{tiger}: TIGER learns hierarchical discrete semantic IDs for items using a vector quantized autoencoder to enhance semantic-level generalization in recommendation.
        \item \textbf{EAGER}~\citep{eager}: EAGER explicitly aligns goals and actions in sequences through a goal-conditioned transformer, improving recommendation accuracy in long-horizon scenarios.
        \item \textbf{HSTU}~\citep{hstu}: HSTU models hybrid short- and long-term user preferences using temporal contrastive learning and dual memory mechanisms for better temporal representation.
        \item \textbf{ActionPiece}~\citep{actionpiece}: ActionPiece segments user behavior sequences into variable-length action pieces to capture high-level user intentions via dynamic granularity modeling.
        \item \textbf{COBRA}~\citep{cobra}: COBRA generates context-aware behavior representations by decomposing and composing user sequences, enabling robust modeling of behavior semantics under sparse supervision.
    \end{itemize} 
\end{enumerate}

\begin{table}
  \setlength{\tabcolsep}{3pt}
  \scriptsize
  \caption{Performance comparison of different methods. The best performance is highlighted in bold while the second best performance is underlined. The last column indicates the improvements over the best baseline models.}
  \label{performence}
  \centering
  \begin{tabular}{llccccccccccc}
    \toprule
    \multirow{2}{*}{Dataset} &\multirow{2}{*}{Metric} &\multicolumn{4}{c}{\textit{Traditional}} &\multicolumn{6}{c}{\textit{Generative}}  &\multirow{2}{*}{Improv.}\\
    \cmidrule(lr){3-6} \cmidrule(lr){7-12} 
    &  &Caser &HGN &SASRec &S3Rec &TIGER &EAGER &HSTU &ActionPiece &COBRA &CAR & \\
    \midrule
    \multirow{4}{*}{Beauty} &Recall@5  &0.0205 &0.0325 &0.0387 &0.0387 &0.0454 &\underline{0.0618} &0.0469 &0.0511 &0.0537 &\textbf{0.0667} &7.93\%    \\
    &Recall@10  &0.0347 &0.0512 &0.0605  &0.0647 &0.0648 &\underline{0.0836} &0.0704 &0.0775 &0.0725 &\textbf{0.0894} &6.94\%    \\
    &NDCG@5  &0.0131 &0.0206 &0.0249  &0.0244 &0.0321 &\underline{0.0451} &0.0314 &0.0340 &0.0395 &\textbf{0.0477} &5.76\%     \\
    &NDCG@10  &0.0176 &0.0266 &0.0318 &0.0327 &0.0384 &\underline{0.0525} &0.0389 &0.0424 &0.0456 &\textbf{0.0549} &4.57\%\\
    \midrule
    \multirow{4}{*}{Sports} &Recall@5  &0.0116 &0.0189 &0.0233  &0.0251 &0.0264 &0.0281 &0.0258 &\underline{0.0316} &0.0305 &\textbf{0.0405} &28.16\%\\
    &Recall@10  &0.0194 &0.0313 &0.0350  &0.0385 &0.0400 &0.0441 &0.0414 &\underline{0.0500} &0.0434 &\textbf{0.0558} &11.60\%\\
    &NDCG@5   &0.0072 &0.0120 &0.0154  &0.0161 &0.0181 &0.0184 &0.0165 &0.0205 &\underline{0.0215} &\textbf{0.0279} &29.77\%\\
    &NDCG@10  &0.0097 &0.0159 &0.0192 &0.0204 &0.0225 &0.0236 &0.0215 &\underline{0.0264} &0.0257 &\textbf{0.0328} &24.24\%\\
    \midrule
    \multirow{4}{*}{Toys} &Recall@5  &0.0166 &0.0321 &0.0463  &0.0443 &0.0521 &0.0584 &- &- &\underline{0.0619} &\textbf{0.0744} &20.19\%\\
    &Recall@10  &0.0270 &0.0497 &0.0675  &0.0700 &0.0712 &0.0714 &- &- &\underline{0.0781} &\textbf{0.0975} &24.84\%\\
    &NDCG@5  &0.0107 &0.0221 &0.0306 &0.0294 &0.0371 &\underline{0.0464} &- &- &0.0462 &\textbf{0.0547} &17.89\%\\
    &NDCG@10  &0.0141 &0.0277 &0.0374 &0.0376 &0.0432 &0.0505 &- &- &\underline{0.0515} &\textbf{0.0621} &20.58\%\\
    
    \bottomrule
  \end{tabular}
\end{table}

\paragraph{Recommendation Performance} We reproduced the experimental results of TIGER~\citep{tiger} as a baseline for comparison, while the results of HSTU are taken from the original ActionPiece paper~\citep{actionpiece}. The remaining baselines are based on either their original publications or publicly available~\footnote{https://github.com/aHuiWang/CIKM2020-S3Rec} implementations provided by Zhou et al.~\citep{s3}.

As shown in Table~\ref{performence}, CAR consistently outperforms existing baselines across all three datasets. Notably, it achieves significant improvements on the Sports and Toys datasets, with relative gains in Recall@5 of 28.16\% and 20.19\%, respectively. Although the improvement on the Beauty dataset is relatively smaller—a 7.93\% increase in Recall@5—CAR still surpasses the second-best baseline, COBRA, by a notable 24.21\% in Recall@5. We hypothesize that the relatively smaller gain on the Beauty dataset can be attributed to the particularly strong performance of EAGER, which raises the baseline and thus narrows CAR’s relative advantage.

\subsection{Ablation Study}

We conducted ablation studies, as shown in Table~\ref{CAR_ablation}, to evaluate the performance advantages of CAR over the standard autoregressive modeling (AR) and to analyze the individual contributions of its core components.

(1) AR vs CAR: While the AR method also represents items using chunks, it adopts the conventional next token prediction as its modeling objective. The results in Table 1 demonstrate that CAR significantly outperforms AR across multiple evaluation metrics. We attribute this performance gain to CAR’s reformulation of the modeling task as next chunk prediction, which effectively decouples the sequential dependency between collaborative and semantic information, thereby enhancing the model’s overall representational capacity.

(2) Ablation of Key Components: We further investigated the impact of two core modules within CAR. ``CAR \textit{w/o} F'' denotes the removal of the Progressive Act-Think Context \textbf{F}usion module, while ``CAR w/o T'' refers to eliminating the \textbf{T}hink loss during training, meaning the model is only supervised to predict the unique ID, without semantic ID supervision. The results show that removing the fusion module causes a significant performance drop, confirming that progressive context integration strengthens the representation of both SIDs and UID. Similarly, excluding the semantic supervision (think loss) also degrades performance, which reflects that explicitly modeling the ``think'' process, reflected in semantic information, is important for improving recommendation accuracy.

\begin{table}
  \caption{Ablation analysis of CAR. ``AR'' refers to standard \textbf{A}uto\textbf{R}egressive modeling. ``CAR \textit{w/o} F'' denotes the removal of the Progressive Act-Think Context \textbf{F}usion module. ``CAR w/o T'' indicates the exclusion of the \textbf{T}hink loss during training. }
  \label{CAR_ablation}
  \centering
  \begin{tabular}{lcccccccc}
    \toprule
    \multirow{3}{*}{method} &\multicolumn{4}{c}{Beauty} &\multicolumn{4}{c}{Sports} \\
    \cmidrule(lr){2-5} \cmidrule(lr){6-9} 
    &Recall &Recall &NDCG &NDCG &Recall &Recall &NDCG &NDCG \\
    &@5 &@10 &@5 &@10 &@5 &@10 &@5 &@10 \\
    \midrule
    AR &0.0373 &0.0521 &0.0269 &0.0321 &0.0164 &0.0105 &0.0251 &0.0133\\
    CAR \textit{w/o} F\&T &0.0610 &0.0824 &0.0442 &0.0510 &0.0343 &0.0479 &0.0244 &0.0287\\
    CAR \textit{w/o} F &0.0628 &0.0853 &0.0454 &0.0526 &0.0356 &0.0497 &0.0250 &0.0296\\
    CAR \textit{w/o} T &0.0644 &0.0872 &0.0467 &0.0540 &0.0389 &0.0541 &0.0273 &0.0321\\
    CAR &\textbf{0.0667} &\textbf{0.0894} &\textbf{0.0477} &\textbf{0.0549} &\textbf{0.0405} &\textbf{0.0558} &\textbf{0.0279} &\textbf{0.0328}\\
    \bottomrule
  \end{tabular}
\end{table}

\subsection{Semantic IDs Generation Analysis}

Locality-Sensitive Hashing (LSH)~\citep{lsh3,lsh2,lsh1} generates semantic IDs through random hash mappings, while the Residual-Quantized Variational Autoencoder (RQ-VAE)~\citep{rqvae} produces semantic IDs via jointly trained deep neural network (DNN) encoders and decoders combined with residual quantizers. To ensure a fair comparison, all methods use semantic embeddings generated by a pretrained Sentence-T5 model~\citep{sentencet5} and uniformly output four layers of IDs, each with a cardinality of 256.

As shown in Table~\ref{code generation}, Res-KMeans consistently outperforms the other two methods, indicating that its parameter-free and non-unique semantic IDs design more effectively preserves the original semantic information of items and better supports CAR in capturing commonalities in semantic associations.

\begin{table}
  \caption{Analysis for different Semantic IDs generation techniques for CAR. We show that Res-KMeans Semantic IDs (SIDs) perform significantly better compared to hashing SIDs and RQ-VAE.}
  \label{code generation}
  \centering
  \begin{tabular}{lcccccccc}
    \toprule
    \multirow{3}{*}{method} &\multicolumn{4}{c}{Beauty} &\multicolumn{4}{c}{Sports} \\
    \cmidrule(lr){2-5} \cmidrule(lr){6-9} 
    &Recall &Recall &NDCG &NDCG &Recall &Recall &NDCG &NDCG \\
    &@5 &@10 &@5 &@10 &@5 &@10 &@5 &@10 \\
    \midrule
    LSH &0.0612 &0.0839 &0.0446 &0.0520 &0.0368 &0.0515 &0.0259 &0.0306 \\
    RQ-VAE &0.0635 &0.0827 &0.0463 &0.0524 &0.0382 &0.0535 &0.0266 &0.0315 \\
    Res-KMeans &\textbf{0.0667} &\textbf{0.0894} &\textbf{0.0477} &\textbf{0.0549} &\textbf{0.0405} &\textbf{0.0558} &\textbf{0.0279} &\textbf{0.0328}\\
    \bottomrule
  \end{tabular}
\end{table}

\subsection{Scaling Effects of SIDs Bit Number in CAR}

We further investigate the impact of the number of bits used in SIDs encoding on CAR’s performance, as reported in Table \ref{semantic bit length}. The hierarchical SIDs structure encodes textual item information from coarse- to fine-grained levels, offering a layered semantic representation that supports more nuanced item prediction. Notably, as the number of SIDs bits increases from one to four, Recall@5 improves by 25.8\% and 30.5\% on the Beauty and Toys datasets, respectively. This trend reveals a clear scaling effect between semantic information and recommendation quality, echoing patterns observed in large language models where richer representations lead to more effective reasoning~\citep{deepseekthink,slow1,slow2}. These findings suggest that CAR is not only capable of capturing fine-grained semantic information, but also preliminarily demonstrates a slow thinking style mechanism—where increasing intermediate computational steps can lead to improved model performance.

\begin{table}
  \caption{Recall and NDCG metrics for different semantic ID bit number.}
  \label{semantic bit length}
  \centering
  \begin{tabular}{lcccccccc}
    \toprule
    \multirow{3}{*}{Length} &\multicolumn{4}{c}{Beauty} &\multicolumn{4}{c}{Toys} \\
    \cmidrule(lr){2-5} \cmidrule(lr){6-9} 
    &Recall &Recall &NDCG &NDCG &Recall &Recall &NDCG &NDCG \\
    &@5 &@10 &@5 &@10 &@5 &@10 &@5 &@10 \\
    \midrule
    1 &0.053 &0.0743 &0.0381 &0.0449 &0.0570 &0.0771 &0.0421 &0.0485\\
    2 &0.0607 &0.0823 &0.0441 &0.0510 &0.0655 &0.0866 &0.0479 &0.0547\\
    3 &0.0635 &0.0856 &0.0459 &0.0530 &0.0699 &0.0916 &0.0512 &0.0582\\
    4 &\textbf{0.0667} &\textbf{0.0894} &\textbf{0.0477} &\textbf{0.0549} &\textbf{0.0744} &\textbf{0.0975} &\textbf{0.0547} &\textbf{0.0621}\\
    \bottomrule
  \end{tabular}
\end{table}

\subsection{Inference Speed Comparison}

We compared the inference speed of standard AutoRegressive (AR) method~\citep{attention,neural} and Chunk Autoregressive (CAR) method on the Toys dataset, with the results shown in Table \ref{inference}. To ensure a fair comparison, the AR method adopts the same GPT-2 architecture and parameter settings as CAR, and item representations are encoded using chunks containing 4-bit SIDs. Due to the need for sequential decoding, AR must rely on beam search during inference, where the search space expands significantly with the number of beams, leading to a notable slowdown in inference speed~\citep{tiger,eager}. In contrast, CAR can decode all tokens in the next chunk in parallel, thereby avoiding the computational overhead introduced by beam search and significantly improving inference efficiency. More importantly, the inference speed of CAR remains relatively stable across different numbers of beams, demonstrating stronger scalability and efficiency advantages.

\begin{table}
  \caption{Inference speed (second per sample) comparison between AR and CAR on Toys dataset.}
  \label{inference}
  \centering
  \begin{tabular}{lccc}
    \toprule
    num\_beams &AR &CAR &Ratio \\
    \midrule
    5 &0.00411 &0.00012 &34\\
    10 &0.00883 &0.00013 &68\\
    15 &0.01449 &0.00013 &111\\
    20 &0.01902 &0.00013 &146\\
    \bottomrule
  \end{tabular}
\end{table}

\section{Conclusion}

In this work, we propose CAR, a novel chunk-level autoregressive framework for generative recommendation that integrates user actions and semantic information through an ``act-with-think'' paradigm. By modeling each item as a SIDs-UID chunk, CAR captures the intertwined nature of collaborative and semantic information, moving beyond the limitations of independent modeling. Our approach not only improves recommendation accuracy, as demonstrated by significant gains on multiple public datasets, but also enhances interpretability by providing a clear semantic information behind each prediction. We believe CAR offers a promising direction for building more explainable and cognitively aligned recommendation systems.

\bibliography{reference}

\end{document}